# Structural and dynamic properties of linker histone H1 binding to DNA

Rolf Dootz,[1] Adriana Cristina Toma[2] and Thomas Pfohl[1,2]*

[1]Max-Planck-Institute for Dynamics and Self-Organization, Bunsenstraße 10,
37073 Göttingen, Germany

[2]Chemistry Department, University of Basel, Klingelbergstrasse 80, 4056 Basel, Switzerland

Corresponding author:   Thomas Pfohl

mail:   thomas.pfohl@unibas.ch

**Found in all eukaryotic cells, linker histones H1 are known to bind to and rearrange nucleosomal linker DNA.** *In vitro*, **the fundamental nature of H1/DNA interactions has attracted wide interest among research communities - for biologists from a chromatin organization deciphering point of view, and for physicists from the study of polyelectrolyte interactions point of view. Hence, H1/DNA binding processes, structural and dynamical information about these self-assemblies is of broad importance. Targeting a quantitative understanding of H1 induced DNA compaction mechanisms our strategy is based on using small angle X-ray microdiffraction in combination with microfluidics. The usage of microfluidic hydrodynamic focusing devices facilitate a microscale control of these self-assembly processes. In addition, the method enables time-resolved access to structure formation** *in situ*, **in particular to transient intermediate states. The observed time dependent structure evolution shows that the interaction of H1 with DNA can be described as a two step process: an initial unspecific binding of H1 to DNA is followed by a rearrangement of molecules within the formed assemblies. The second step is most likely induced by interactions between the charged side chains of the protein and DNA. This leads to an increase in lattice spacing within the DNA/protein assembly and induces a decrease in the correlation length of the mesophases, probably due to a local bending of the DNA.**

\body

The linker-histone family is a heterogeneous family of highly tissue-specific, basic proteins, which exhibit significant variations in sequence (1,2). However, most eukaryotic linker-histones H1 share a similar, tripartite structure consisting of a globular domain flanked by two lysine-rich tails, a shorter amino-terminal domain (N-tail) and a longer, carboxyl-terminal one (C-tail) (3). Linker-histones H1 are known to attach close to the entry and exit sites of linker-DNA on the nucleosome core bringing together two linker-DNA segments (4-6). Studies have shown that the globular domain is responsible for this positioning (7,8). The globular domain with a diameter of 2.9 nm is the only domain that is folded in solution exhibiting three $\alpha$-helices (9-11). Its most noticeable structural features are the two DNA-binding sites situated on opposite sides of the molecule (12), which with the help of the C-terminal domain render H1-chromatin binding highly dynamic (13,14).

Despite positioning along the nucleosome core particle, it is not the globular domain but rather the highly positively charged C-tail that imparts to linker-histones their unique ability to bind to linker-DNA through non-specific electrostatic interactions (6,15,16). *In vivo*, the absence of C-tails leads to greatly reduced chromatin binding (17). Binding of linker-histones to the linker-DNA facilitates the shift of chromatin structure towards more condensed, higher order forms (18) Although a chromatin fiber lacking linker-histones is able to fold to a certain extent (19), there is abundant evidence that the highly ordered chromatin compaction of the 30 nm fiber is only attained in the presence of linker-histones (5,18,20). *In vivo* studies on H1 depleted chicken cells have shown an altered chromatin structured as well as chromosomal aberrations and increased DNA damage during replication, suggesting that H1 was involved in transcription regulation (21). Linker-histones help to select a specific folding state from among the set of compact states reached in its absence by contributing to the free energy of chromatin folding (22). This suggests that linker-histones are of central importance in genome organization and regulation.

Since the linker histone's position on the nucleosome is still a matter of debate, understanding more closely its interaction with DNA upon condensing it is therefore essential for understanding the role H1 plays in gene regulation and chromatin structure. It has been argued that H1/DNA assemblies are an excellent model system for studying aspects of the interaction of H1 with chromatin. H1/DNA interaction *in vitro* (15, 23-31) was reported to be dependent on the ionic strength (23, 31-33), suggesting an interaction governed by electrostatics. At high salt concentrations the screening of interactions was so significant that assembling was no longer reported (34), denoting a rather weak binding in comparison to protamine-DNA assemblies that can still be formed even at monovalent salt concentrations as high as 1.3M (35). Thus, reinforcing the belief that H1 is a highly mobile chromatin component compared to the germinal chromatin which is made mechanically stable and transcriptionally inactive by strong protamine binding. In a more general context, the H1/DNA system can be regarded as a model for non-specific DNA-protein interactions. Herein, using the innovative junction between X-ray scattering and microfluidics, we probed linker histone/DNA interaction dynamics and structure formation in real-time. A microfluidic channel with a cross geometry device provides means to control mixing of H1 and DNA uncovers a two-step assembly process. The non-specific binding of linker histones to DNA lead to the formation of primary assemblies with a columnar hexagonal structure governed by electrostatic interactions. As the binding reaction evolves and the concentration of H1 histones increases, the columnar structure rearrange to less organized assemblies with smaller correlation lengths. We attribute the latter observation to an additional bending of DNA molecules induced by the interaction of the linker histones tails with DNA.

**Results and discussion.**

Aside from advantages such as reduced sample volumes and the possibility of high throughput and parallel operations, microfluidic techniques are particularly useful for investigations of biomaterials. Importantly the combination of using microfluidics and X-rays, significantly reduces radiation induced damage of the sample - an important experimental consideration given the damage X-rays cause to protein solutions (36). The hydrodynamic focusing device used here consists of two perpendicular microchannels in the form of a cross with three inlets and one outlet. A semi-diluted aqueous DNA solution is injected in the reaction channel and hydrodynamically focused by two side streams of aqueous H1 solutions. Figure 1 gives a schematic representation of the experimental setup.

Laminar flow conditions due to microscale channel dimensions force mixing to occur purely by diffusion. Following the confluence of the microchannels, H1 molecules diffuse into the DNA stream establishing stable concentration gradients. Mixing and concentration distributions in the reaction channel can be adjusted by controlling main and side channel velocities. Flow velocities are chosen such that concentration gradients of reactants extend along the measurable length of the device. It follows that the composition of the formed assemblies varies at every accessible point along the reaction channel. This allows the access to the dynamics of H1/DNA structure formation *in situ*, in particular to transient intermediate states. Thus, it is essential to first quantify concentrations and therefore the composition of the aggregates. This is achieved by a detailed comparison of microscopic images and finite element simulations of flow conditions in the microfluidic device.

**Quantifying local conditions in microfluidic devices.** Polarized light microscopy has been used widely to study DNA assemblies, which are known to be highly birefringent (37-39). The birefringence signal is increased upon self-assembly of DNA aggregates under conditions of alignment and elongation which are imposed by using microfluidic devices. Concurrently, combining microfluidics with polarized light microscopy provides a fast and easy access to direct imaging of H1 induced DNA compaction. Data are acquired in the reaction channel (main channel) at three different flow velocities $u_{DNA} = 60$, 150, and 600 µm·s$^{-1}$. The flow velocities in the side channels are varied such that a flow velocity ratio $u_{H1}/u_{DNA} = 1$ is maintained.

Finite element simulations of the flow profiles inside the microchannels are performed and compared to experiments in order to elucidate the experimental parameters. In Figure 2a, simulation data are shown for $u_{DNA} = 600$ µm·s$^{-1}$ and contrasted to corresponding experimental results. Owing to the symmetry of the microfluidic device in two dimensions, it is sufficient to simulate half of the device. Following the intersection of the two fluid flows, H1 molecules diffuse into the DNA stream and the H1/DNA interactions can be observed along the reaction channel. The optically birefringent pattern reflects that DNA chains in the assemblies are orientationally ordered due to the superimposed flow. For direct comparison with experimental results, the modeled velocity profile (white arrows in Figure 2a) in the hydrodynamic focusing device is overlaid to the recorded birefringence image (Figure 2a bottom). The strong increase in local viscosity connected to the compaction reaction can be exploited to visualize H1 DNA assemblies in the simulations. Insignificant deviations of simulation results from the experimentally recorded shape are observed in the crossing area, where the center stream is slightly expanding into the side channels. These deviations result from the fact that simulations are performed in 2D whereas the experimental system is affected by additional walls at the top and the bottom of the device (40). Apart from this detail, experiments and simulations show good agreement. The result of the corresponding simulations over the whole range of the device is given in Figure 2b.

From simulations, local experimental parameters such as flow velocities and concentrations can be obtained at each position. Local concentrations are translated into assembly compositions given in terms of the relative charge ratio $N/P$. $N$ is the total number of positive amine charges of H1 and $P$ is the total number of negatively charged DNA phosphate groups. In Figure 2c, $N/P$ ratios are plotted for three different flow velocities as a function of the position $x$ along the center of the outlet channel ($y = 0$). Flow velocities result in final charge ratios at the furthest measurable point of the device ($x \approx 12$ mm) of $N/P <$ 2.4, 3.3, and 5.1 for $u_{DNA} = $ 60, 150, and 600 $\mu m \cdot s^{-1}$, respectively. Local flow velocities can be used to translate positional changes along each streamline into corresponding reaction time coordinates $t$. Figure 2d shows the $t$ dependence of $N/P$ represented as well for three different flow velocities. For larger $t$, $N/P$ ratios unite into o a single curve. Deviations at initial time states reflect differences in the flow velocity depending strain rate $\dot{\varepsilon} = \partial u / \partial y$ (41). Once the information concerning the local concentrations and hence assemblies' compositions is at hand, it is possible to analyze H1/DNA structure formation in detail.

**Small angle X-ray diffraction of H1/DNA assemblies.** Spatially resolved small angle X-ray (micro-)diffraction is employed as the principle method of analysis to access relevant molecular length scales for the study of biomolecular interactions. Data are obtained at different $x$-positions along the main channel for all three different flow velocities. In Figure 3, characteristic 2D diffraction images are shown. The observed alignment is due to the elongational flow at the confluence of the solution streams. Structural information can be obtained by analyzing the radial integrated intensity profiles. The diffraction intensity is plotted as a function of the scattering vector $q$, which is inversely proportional to the periodic distance between reticular planes $d$ (42). Since DNA has a significantly higher electron density than H1 proteins, the DNA self-assembly promotes or leads to the formation of mesophases that dominate the scattering profile. At positions close to the middle of the cross channels, the diffraction intensity curve is composed of a broad peak with a shoulder in the low $q$ values region. The best decomposition of this broad peak is successfully made by fitting two Lorentzian functions yielding peak positions $q_1$ and $q_2$. In Figure 3, this is shown for the scattered intensity at $x = 100$ μm.

In Figure 4a, the dependence of the peak positions $q_1$ and $q_2$ on the channel position $x$ is given for the data set recorded at $u_{DNA} = 150$ $\mu m \cdot s^{-1}$. To elucidate their dependence on assembly composition, it is reasonable to plot quantities of interest versus $N/P$ obtained from simulations. This is shown for $q_1$ and $q_2$ in Figure 4b.

Plotting quantities extracted from X-ray diffraction data obtained at different flow velocities against $N/P$ allows for superposition of all data onto a single plot showing their dependence on composition. In Figure 5a this is demonstrated for peak positions $q_1$ (lower curve) and $q_2$ (upper curve) measured along the streamline in the center of the reaction channel ($y = 0$). The three data sets obtained at $u_{DNA} = 60$, 150, and 600 µm·s$^{-1}$ show excellent agreement with deviations between different data sets of less than 0.01 nm$^{-1}$. Local $N/P$ ratios are highly dependent on the diffusion of H1 molecules. Accordingly, the fact that data obtained at different flow velocities are in good agreement and collapse onto single curves establishes the validity of the experimental method and the high degree of consistency between experiments and simulations.

At low $N/P$ ratios, peak positions of $q_1 = 1.76$ nm$^{-1}$ and $q_2 = 1.90$ nm$^{-1}$ are observed. With increasing $N/P$, $q_1$ and $q_2$ are simultaneously shifted toward lower $q$ values with minima at $N/P \approx 0.2$, of $q_1 = 1.73$ nm$^{-1}$ and $q_2 = 1.88$ nm$^{-1}$, respectively. Following, the peak position $q_1$ increases monotonically moving towards higher $q$ values whereas $q_2$ is levels off at $q_2 = 1.90$ nm$^{-1}$. Eventually, for all studied velocities, for $N/P > 1.8$ (corresponding to $x > 2200$ µm) in Figure 5a, the peak at $q_2$ disappears leaving a single peak at $q_1$. Associated with the disappearance of the peak at $q_2$, the remaining peak $q_1$ shows a $q_{max}$ position at 1.78 nm$^{-1}$. Increasing H1 concentration even further (i.e increasing the charge ration N/P) results in smaller $q$ values yielding $q_1 = 1.72$ nm$^{-1}$ at the furthermost observable position $x = 12000$ µm along the reaction channel ($N/P \approx 3.3$, Figure 5) for a flow velocity of 60 µm·s$^{-1}$.

**Structure of H1/DNA mesophases.** Owing to the absence of higher order peaks, the detailed structure of formed H1/DNA mesophases cannot be ruled out. This is mainly due to the fact that spatial constrains of the beam-line limited the observable $q$-range to $q < 2.67$ nm$^{-1}$. Furthermore, correlation lengths of formed H1/DNA aggregates are on the order of 10-70 nm. For systems with such a reduced long-range ordering, scattering peaks of a structure with square symmetry is expected at a position $q_1 \cdot \sqrt{2} \approx 2.43$ nm$^{-1}$, which is well situated in the accessible $q$-range and should be therefore observable. Furthermore, minima of the form factor of the globular domain, which could account for an absence of the (110) peak, are situated at 1.55 and 2.66 nm$^{-1}$ and are therefore not expected to be of influence. Accordingly, from an absence of peaks at this position, the in-plane structure of the mesophase exhibiting the peak at $q_1$ can be ruled out to be most likely a hexagonal one. Unfortunately, it is not possible to narrow down the structure of the mesophase exhibiting the peak at $q_2$. Assuming hexagonal ordering, the lattice spacing $d$ can be calculated according to the following relation: $d = 4\pi/\sqrt{3}q$. In Figure 5b, lattice spacings $d_1$ and $d_2$ – corresponding to $q_1$ and $q_2$ – are given and their dependence on H1/DNA assembly composition is shown in terms of the charge ratio $N/P$. Observed lattice spacings are in the range of 3.8-4.2 nm.

The particular three domain structure of the linker histone induces a complex interaction with DNA leading to the formation of unique structures. Although unique, the DNA molecules within these structures are organized in a columnar hexagonal phase as it was shown in earlier studies using polarizing microscopy for investigating DNA dense phases induced by other polycations (45). Previous work using electron microscopy grids to visualize H1/DNA complexes on an electron microscopy grid revealed that H1 molecules are sandwiched between two DNA helices forming a tram-track like pattern (12, 24, 25, 29) with a diameter of 3.8 nm (23), agreeing with interhelical values $d_2$ plotted in Figure 5b. However, interhelical distances within complexes formed with poly-L lysine (PLL) at a N/P ratio of 1.7 and no added salt, were found close to 2.7 nm (42), thus below our experimental values. Although the inner organization of the helices with the linear PLL is the same (hexagonal packing) the higher lattice values we found are clearly due to the structured globular domain of the linker histones. In order to enable comparability with results in literature, it is useful to translate $N/P$ into mass fraction $w/w$ of H1 to DNA. The X-ray data presented in Figure 5b exhibit a maximal lattice spacing of both coexisting phases at $N/P \approx 0.2$ ($w/w \approx 0.2$). This is in line with the fact that for linear DNA molecules and low-salt conditions small amounts of H1, $w/w \approx 0.15$, produce complete incorporation of all DNA molecules into extremely large aggregates (26).

**Dynamics of H1/DNA assemblies structure formation.** Figure 3 shows the existence of two overlapping diffraction peaks. The evolution of these peaks at different positions along the reaction channel is an evidence of a two-step process. At first, H1 binds electrostatically to DNA and the formed assemblies give rise to the diffraction peak at $q_2$. A closer study of this diffraction peak reveals columnar hexagonal packing with DNA helices linked together by H1 molecules. The second step of the process is most probably due to the successive rearrangement of molecules in the formed assemblies which result in a structure yielding the peak at $q_1$. Infrared (IR) spectroscopy studies have shown changes in the structure of the C terminal domain of H1 upon DNA binding (46). In our experiments, this conformational transition can be monitored in terms of the intensity ratio $I_2/I_1$ of the two Bragg reflections (Figure 5c). A coexistence regime occurs over a relatively wide range of $N/P$ ratios. It is characterized by an overlapping of the two relatively sharp peaks at $q_1$ and $q_2$. With increasing $N/P$, $I_2/I_1$ is gradually reduced reaching zero at $N/P \approx 1.8$. This is the assembly composition at which the peak at $q_2$ is completely lost. The scattering is further characterized by a single peak at $q_1$ that both shifts to lower $q$ values and broadens in $q$ with a further increase of $N/P$.

$N/P \approx 1.8$ corresponds to 15 base pairs (bp) of DNA per H1 molecule. This result is in excellent agreement with sedimentation titration binding data that reported a binding density of one H1 molecule per 10-13 bp ($N/P \approx 2.5$-1.9) (47) - a value relatively independent of salt concentration in the range of $c_{salt}$ = 14 mM-350 Mm (31, 47). Moreover, our result is also consistent with nuclease digestion studies of chromatin which have shown that each linker-histone protects approximately 10 bp from each end of the chromatosomal DNA (20, 32, 48). The experimental setup we used did not allow allow for a complete distinction between reaction time and composition dependent effects. In Figure 2d, the dependence of $N/P$ on the reaction time $t$ is given for all three flow velocities. For flow velocities of $u$ = 150 $\mu m \cdot s^{-1}$ and 600 $\mu m \cdot s^{-1}$, $N/P \approx 1.8$ is reached at $t \approx 2.5$ s. For $u$ = 60 $\mu m \cdot s^{-1}$, this assembly composition is only reached at $t \approx 4.1$ s. These deviations are due to several factors such as the complex interaction of the flow fields influenced by the local viscosities and diffusion. However, although the reaction time is almost doubled for $u$ = 60 $\mu m \cdot s^{-1}$, a vanishing of the peak at $q_2$ is only observed for $t \geq 4.1$ s. This indicates that the compaction mechanism of H1 and DNA is rather diffusion limited and conforms with observations of similar sized dendrimer/DNA interactions.

**Domain sizes of H1/DNA assemblies.** In addition to peak positions, average domain sizes of H1/DNA assemblies can be determined from the full width at half maximum $\Delta q$ of the reflections at $q_1$ and $q_2$. The domain size corresponds to a typical correlation length $L_C = 2\pi/\Delta q$. To ensure comparability, it is useful to analyze $q/\Delta q$, which corresponds to the correlation lengths given in terms of the lattice spacing, $q/\Delta q = L_C/d$. The $N/P$-dependence of $q_1/\Delta q_1$ and $q_2/\Delta q_2$ are shown in Figure 5d. A clear dependence on the flow velocity of both $q_1/\Delta q_1$ and $q_2/\Delta q_2$ is evident. At low $N/P$ ratios in the coexistence region, $q_2/\Delta q_2$ is significantly higher than $q_1/\Delta q_1$ exhibiting values of 15-23 ($L_{C2} \approx 48$-73 nm) and 5-9 ($L_{C1} \approx 19$-33 nm), respectively. With increasing $N/P$, $q_2/\Delta q_2$ shows a strong decrease starting around the charge neutral point characterized by $N/P = 1$. For all three flow velocities, maximal values $q_1/\Delta q_1$ = 6, 7, and 9, respectively, are found at $N/P \approx 1.8$ when the feature at $q_2$ is disappeared. This finding is consistent with the evolution of the ratio of intensity $I_2/I_1$ shown in Figure 5c. Parallel to the observed shift in peak position $q_1$ to smaller $q$ values with further increasing H1 concentration, domain sizes decrease to about $4 \cdot d$ at the furthest recorded assembly composition ($N/P \approx 3.3$).

The ratio $q/\Delta q$ allows for comparison of H1/DNA domain sizes to values obtained in dendrimer/DNA assemblies. Compared to linker-histones, PPI generation 4 and PAMAM generation 3 dendrimers have a similar size and charge. These DNA assemblies have similar lattice spacings ($d_{PPI4}$ = 3.1-3.6 nm, $d_{PAMAM3}$ = 3.8-4.3 nm, $d_{H1}$ = 3.8-4.2 nm). At comparable strain rates ($\dot{\varepsilon}_{max} \approx 1-2 \ s^{-1}$) and charge ratios ($N/P \approx 1$), PPI 4/DNA and PAMAM 3/DNA assemblies exhibit domain sizes of approximately 23·$d$ and 27·$d$, respectively. These values are comparatively close to those of $q_2/\Delta q_2$ and differ significantly from $q_1/\Delta q_1$.

**Two mesophases having different linker histone-dependent compaction and structures.**
The X-ray diffraction patterns show that at low charge ratios two mesophases coexist within the H1/DNA assemblies. Since both phases experience identical experimental conditions ($N/P$, strain rate,…) the differences in correlation lengths of $L_{C2} \approx 2.5$-$3 \cdot L_{C1}$ suggest that the transition from an ordered mesophase ($L_{C2}$) corresponding to the columnar hexagonal organization of helices immediately after binding H1, to a less organized structure ($L_{C1}$) is mediated by the rearrangement of the histone tails. For a clearer illustration of the two-step model, a schematic representation is shown in Figure 6. As presented in Figure 5d the correlation length of domains with extended tails ($L_{C2}$) is rapidly reduced after $N/P \approx 1.1$ (28 bp of DNA per each H1 molecule) implying that the tails distort the order of the columnar phase and bend the DNA. Several studies have previously suggested that chromatosomal linker DNA is bent by the C-terminal domain of H1 forming a stem-like structure (49-52). This structure has also been implicated in the formation of the 30 nm chromatin fiber (53).

**Conclusion.**
Based on the combination of microfluidics technology with X-ray microdiffraction we demonstrate that the interactions of H1 with DNA follow a two-step dynamic process. The efficiency of DNA compaction by H1 is influenced by multiple factors including flow velocities, diffusion and viscosity. We show that the first binding step is primarily due to electrostatic interactions between the DNA and the linker histone. Our results suggest that the organization of this phase is columnar hexagonal and is composed of domains with a long correlation length. Further on, due to the most certain rearrangement of histone tails within this dense phase, a loss of ordering is observed. Thus, domains with shorter correlation lengths are revealed which can be attributed to an additional bending of the DNA.

A potential direction for future studies involves monitoring of nucleosome core particle arrays/H1 assemblies dynamic formation and structure evolution in microflow. These assemblies may shed light on the 30 nm fiber formation in real-time. Our approach may be generalized and used to access additional relevant biophysical problems of chromatin compaction and decompaction in a single microfluidic experiment.

**Materials and methods.**

**Materials.** Calf thymus linker-histone H1 (isolated lysine rich fraction (54)) and lyophilized highly-polymerized calf-thymus DNA were purchased from Sigma-Aldrich GmbH, Taufkirchen, Germany. The average DNA chain length was determined from its molecular weight on an agarose gel, and a length of 6 μm was estimated. Both components were solubilized in 18.2 MΩcm water (Millipore GmbH, Schwalbach, Germany) to final concentrations of $c_{H1} = 10$ mg·mL$^{-1}$ and $c_{DNA} = 2.5$ mg·mL$^{-1}$ respectively. The pH of both solutions was adjusted by adding HCl$_{aq}$ and NaOH$_{aq}$, respectively. At physiological pH conditions, H1 possesses a molecular weight of $M_w = 21.5$ kDa and 55 positive charges (31), whereas DNA molecules carry two negative charges per base pair.

**Microfluidic devices.** X-ray compatible Kapton-Steel-Kapton microfluidic devices have been fabricated as described elsewhere (55). Briefly, the microchannel structure is spark eroded in a thin stainless steel plate resulting in a microchannel structure, which is open on both sides. The thickness of the plate defines the height of the microchannels. Adhesive Kapton foils coated with a poly(siloxane) layer (thickness 53μm, Dr.Müller GmbH, Allingen, Germany) are placed on both sides of the steel plate, such that the foils seal the device and serve as X-ray transparent windows to the microchannel. Four holes are punched into the bottom Kapton foil fitting the channel ends of the steel plate and serving as inlets of the microfluidic device. Using a thin double sided sticking tape with cavities at the inlet positions and the measuring area, the microfluidic device is mounted on a poly(methylmethacrylate) (PMMA) slab assisting the connection to the fluid pumping system. The center region of the PMMA support is milled out to provide an undisturbed pathway for X-ray beams. The connection to the pumping system is established by Teflon tubing implemented into male nuts (ProLiquid, Überlingen, Germany) that are screwed in the sockets of the PMMA support. The channels we used had a width of 100-150μm and a depth ranging from 200-300μm.

**Microfocused small-angle X-ray measurements.** SAXS experiments were performed at Beamline ID10b at the European Synchrotron Radiation Facility in Grenoble, France. 2D scattering patterns were collected using a CCD detector with a fluorescent screen. Beryllium compound refractive lenses (CRL) focused the X-ray beam of 8 keV ($\lambda = 0.155$ nm) down to a diameter of 20 µm. The microchannel device was loaded on an *x-y* stage, to probe specific positions using the microfocused X-ray beam. The positional accuracy of absolute coordinates in the microdevice was on the order of the beam size. All CCD images were taken at ambient temperature with exposure times of 30 s per position and azimuthally averaged to produce 1D intensity profiles $I(q)$. Using a Lorentzian fit the peak positions and $\Delta q$ in $I(q)$ are determined.

**Finite Element Modeling.** FEMLab software (Comsol, Inc., Burlington, MA) was used to perform finite element modeling simulations of conditions within the microchannel device. The incompressible Navier-Stokes equation was solved in 2D using about 20000 elements to obtain a solution for the diffusive mixing in the microflow. The velocity fields (and corresponding strain rates per position) and concentration profiles were subsequently calculated. The viscosity of the formed H1/DNA assemblies' $\eta_{complex}$ and the diffusion constant $D_{H1}$ were used to match the experimentally recorded shape of the hydrodynamically focused DNA stream and of the formed H1/DNA aggregates. All other parameters such as channel geometry, flow rates, and the viscosity of the DNA solution are known. For each flow velocity, two finite element simulations are performed. In order to accurately determine relevant fit parameters, physical conditions in the microchannel device are first simulated with high precision (i.e. high number of finite elements) for a close-up region around the confluence area ($x = -200$-$500$ µm, Figure 2a). Independent of the flow velocity and throughout all simulations the viscosity of H1/DNA aggregates was fitted to $\eta_{complex} \approx 3 \cdot 10^3 \cdot \eta_{water} = 2.7$ Pa·s. This result is on the same order of magnitude as results known from other polymer hydrogels (40, 56, 57). A diffusion constant of $D_{H1} \approx 2 \cdot 10^{-10}$ m²·s⁻¹ was found, which is close to the result one obtains from the Stokes-Einstein relation ($r_{globular\ domain} \approx 1.5$ nm, $D_{SE} \approx 1.7 \cdot 10^{-10}$ m²·s⁻¹) under purely aqueous conditions. To describe physical conditions at positions further down the reaction channel, a second simulation extending over the whole length of the device ($x = -200$ - $12000$ µm, Figure 2b) has been performed using fit parameter values determined in the first set of simulations.

**Acknowledgements.**
We gratefully acknowledge fruitful discussions with Heather Evans, Stephan Herminghaus, Sarah Köster, Alexander Otten, Bernd Struth, Oleg Konovalov, Anatoly Snigirev, Dagmar Steinhauser, and Sravanti Uppaluri. We thank Udo Krafft for excellent technical assistance. We acknowledge the ESRF for provision of synchrotron radiation facilities at beamline

ID10b. This work was supported by the DFG within the Emmy-Noether-Program (PF 375/2) and SFB 755 "Nanoscale photonic imaging".

**Figure captions.**

**Figure 1.** Schematic representation of the experimental X ray setup. I) Syringe pumps with aqueous solutions of DNA and H1. II) Kapton microfluidic device. III) Microfocused X-ray beam. IV) X-ray diffraction pattern.

**Figure 2.** Real-time monitoring of linker-histone H1 induced DNA compaction in a hydrodynamic focusing device. **(a)** Simulation results (top) and birefringence data (bottom) close to the confluence region are contrasted ($u_{DNA}$ = 600 µm·s$^{-1}$). The product of the assembly reaction appears in the diffusion cone of side and main stream components due to its highly increased viscosity. **(b)** Simulated H1 concentration profile of the whole device ($x$ = -200 - 12000 µm). Dependence of the *N/P* ratio on the position along $x$ **(c)** and the time $t$ **(d)** in the middle of the outlet channel ($y = 0$).

**Figure 3.** Representative 2D X-ray diffraction images (right) obtained at $u_{DNA}$ = 150 µm·s$^{-1}$ in the middle of the outlet channel ($y = 0$) at different positions $x$ and the extracted, radially averaged $q$-dependence of scattering intensities (left). $I(q)$-plots are offset for clarity.

**Figure 4.** Dependence of peak positions $q_1$ and $q_2$ on the position $x$ along the outlet channel **(a)**, and on the *N/P* ratio **(b)** shown exemplarily for the data set recorded at a flow velocity of $u_{DNA}$ = 150 µm·s$^{-1}$.

**Figure 5.** Dependence of the peak position **(a),** the lattice spacing **(b)**, the intensity ratio of the two peaks **(c)**, and the correlation length given in terms of lattice spacings **(d)** on the local *N/P* ratio.

**Figure 6.** Schematic representation of the H1/DNA interaction mechanism. In a first step, H1 molecules bind unspecifically to DNA with extended tails. In a second step, H1 tails fold upon interaction with DNA, distorting and bending thereby the DNA structure.

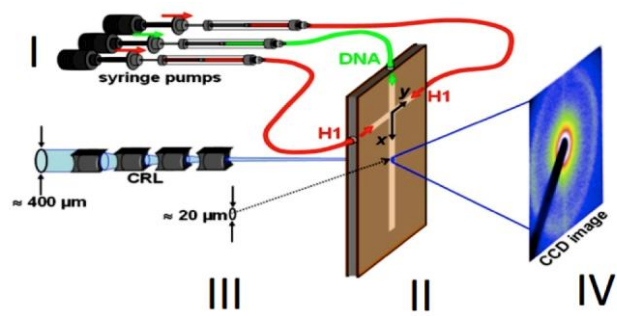

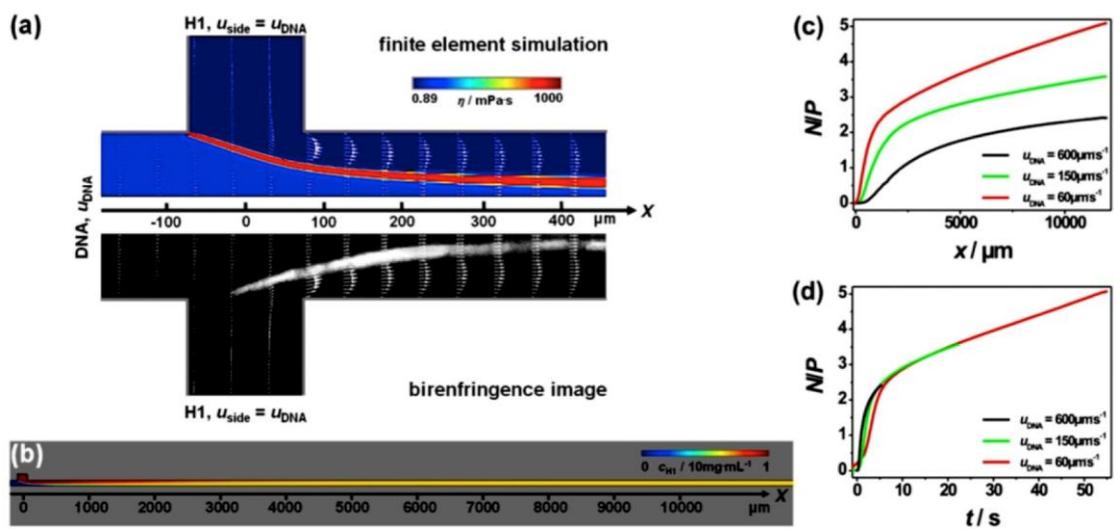

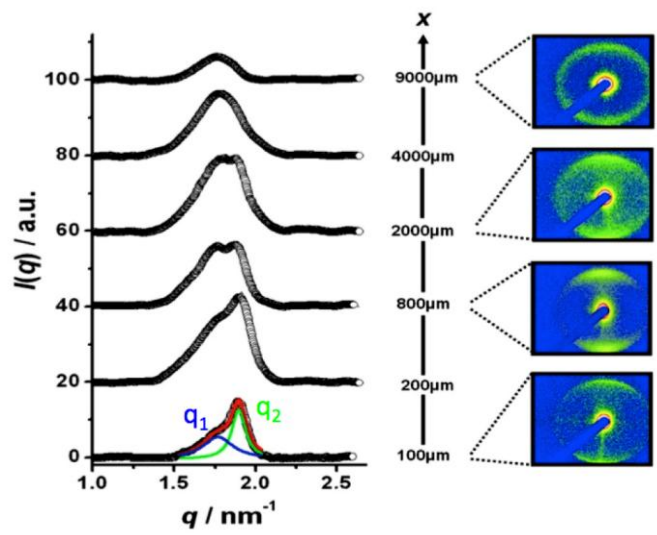

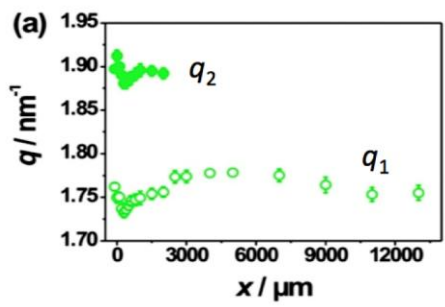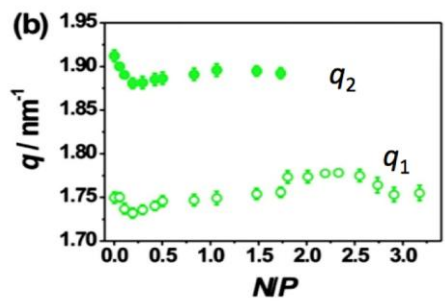

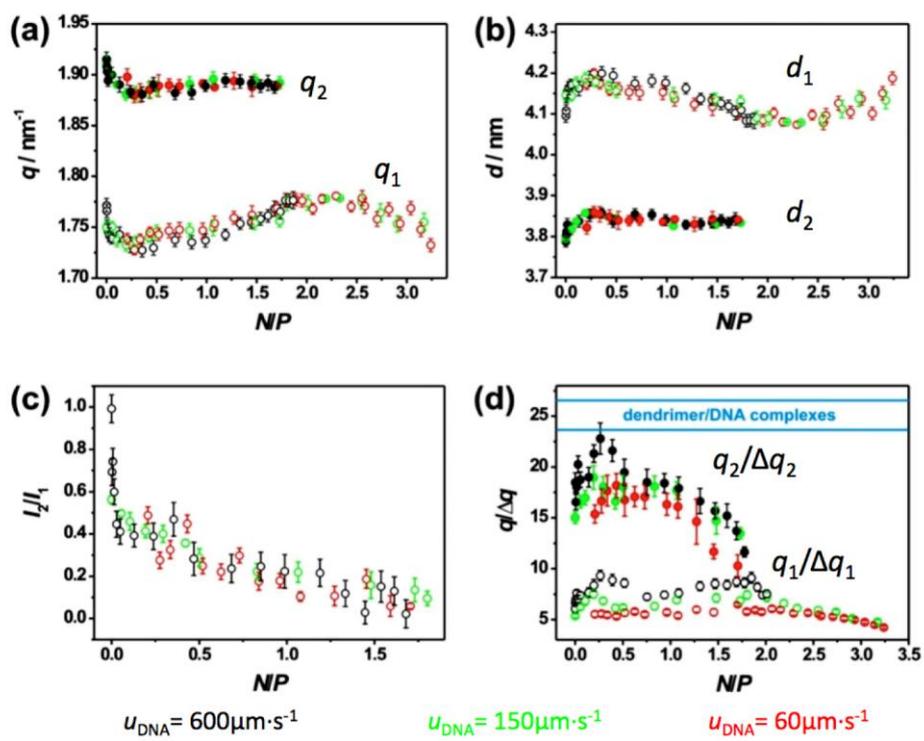

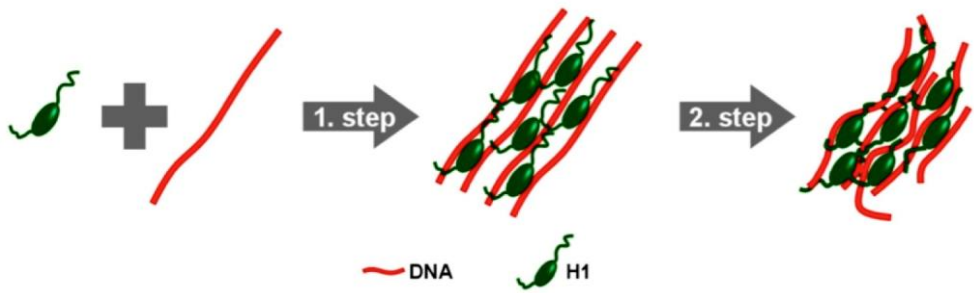